\documentclass[conference]{IEEEtran}
\IEEEoverridecommandlockouts

\usepackage{cite}
\usepackage{amsmath,amssymb,amsfonts}
\usepackage{algorithmic}
\usepackage{graphicx}
\usepackage{textcomp}
\usepackage{xcolor}
\def\BibTeX{{\rm B\kern-.05em{\sc i\kern-.025em b}\kern-.08em
    T\kern-.1667em\lower.7ex\hbox{E}\kern-.125emX}}

\usepackage{ntheorem}
\theoremstyle{plain}
\theoremseparator{.}
\theoremheaderfont{\bfseries\upshape}
\theorembodyfont{\itshape}

\usepackage{algorithm}

\newtheorem{theorem}{Theorem}
\newtheorem{example}{Example}
\newtheorem{lemma}{Lemma}
\newtheorem{construction}{Construction}

\newtheorem{definition}{Definition}

\newcommand{\code}{\ttfamily\bfseries}

\usepackage{pgf}
\usepackage{tikz}
\usetikzlibrary{arrows, automata, positioning, calc, shapes}

\begin{document}

\title{Coding schemes for locally balanced constraints}

\author{
\textbf{Chen Wang},\!\IEEEauthorrefmark{1}
\textbf{Ziyang Lu},\!\IEEEauthorrefmark{1}
\textbf{Zhaojun Lan},\!\IEEEauthorrefmark{2}
\textbf{Gennian Ge},\!\IEEEauthorrefmark{2}
and \textbf{Yiwei Zhang}\IEEEauthorrefmark{1}\\

\IEEEauthorblockA{\IEEEauthorrefmark{1}Key Laboratory of Cryptologic Technology and Information Security of Ministry of Education, \\ School of Cyber Science and Technology, Shandong University, Qingdao, Shandong, 266237, China\\}
\IEEEauthorblockA{\IEEEauthorrefmark{2}School of Mathematical Sciences, Capital Normal University, Beijing 100048, China\\}
\begin{small}{\code \{cwang2021, zylu\}@mail.sdu.edu.cn, zjlan@cnu.edu.cn, gnge@zju.edu.cn, ywzhang@sdu.edu.cn}\end{small}
\thanks{Research supported by National Key Research and Development Program
of China under Grant Nos. 2020YFA0712100, 2021YFA1001000 and 2018YFA0704703, National Natural Science Foundation of China under Grant Nos. 12001323 and 11971325, Shandong Provincial Natural Science Foundation under Grant No. ZR2021YQ46, and Beijing Scholars Program.}}

\maketitle

\begin{abstract}
Motivated by applications in DNA-based storage, we study explicit encoding and decoding schemes of binary strings satisfying locally balanced constraints, where the $(\ell,\delta)$-locally balanced constraint requires that the weight of any consecutive substring of length $\ell$ is between $\frac{\ell}{2}-\delta$ and $\frac{\ell}{2}+\delta$. In this paper we present coding schemes for the strongly locally balanced constraints and the locally balanced constraints, respectively. Moreover, we introduce an additional result on the linear recurrence formula of the number of binary strings which are $(6,1)$-locally balanced, as a further attempt to both capacity characterization and new coding strategies for locally balanced constraints.
\end{abstract}

\section{Introduction}\label{Sec:intro}
With the rapid development of DNA synthesis and DNA sequencing technology,
DNA storage is becoming a promising direction for future data storage \cite{Potomac}.
Error-correcting codes for DNA storage differ from traditional coding theory for communications or current storage mediums in many aspects.
In particular, specific synthesis and sequencing methods and the biochemical properties of DNA strings bring in many additional constraints on the codewords.
Most common constraints are either {\it run-length-limited (RLL) constraints} which limit the length of consecutive repeated symbols, or {\it global GC-content constraints} which require that the percentage of guanine (G) and cytosine (C) in a DNA string must be bounded by a given interval.
Constructing codes with RLL constraints and/or global GC-content constraints is not a new topic and it belongs to the well-established field of constrained coding theory, see for example, \cite{Knuth,MarcusBook}.
Constrained coding problems also have applications in fields other than DNA storage.
For example, constrained codes are extremely useful in communication and storage systems which require simultaneous energy and information transfer \cite{Varshney-ISIT-08}.

In addition to the run-length and global GC-content constraints, during DNA storage some PCR amplification techniques  require local GC-content constraints on DNA strings \cite{Benita-NAR-03}.
That is, within each consecutive substring of a given length, the percentage of guanine and cytosine is also bounded.
Motivated by this application, Gabrys et al. \cite{Gabrys-ISIT-LBC-20} considered the binary case and proposed the so-called {\it locally balanced constraints} and {\it strongly locally balanced constraints}, where the $(\ell,\delta)$-locally balanced constraint requires that in a binary string of length $n$ any consecutive substring of length $\ell$ should have weight between $\frac{\ell}{2}-\delta$ and $\frac{\ell}{2}+\delta$.
In \cite{Gabrys-ISIT-LBC-20} the authors studied the capacity of binary strings meeting such constraints via a spectral graph theory method.
While some capacity results were given in \cite{Gabrys-ISIT-LBC-20}, explicit constructions of codes achieving the capacity and related encoding and decoding algorithms were not considered.
The explicit coding schemes, when $\ell$ and $\delta$ are fixed constants, are the main objective of this paper.

It should be noted that another related work by Nguyen et al. \cite{Nguyen-IT-slidingwindow-21} considered the coding schemes for locally balanced constraints (in the name of sliding window-constrained codes).
However, their results hold only when $\ell=\Omega(\log n)$ and $\delta=\Theta(\ell)$ and their sequence replacing techniques cannot be easily generalized to the case when $\ell$ and $\delta$ are fixed constants.

The rest of this paper is organized as follows. In Section \ref{sec:pre} we introduce the definitions and related results. Two coding schemes are introduced in Section \ref{sec:SLB} for the strongly locally balanced constraints and another scheme is introduced in Section \ref{sec:LB} for the locally balanced constraints. As a further attempt to the capacity characterization and new coding strategies, an additional counting result specifically targeted at the $(6,1)$-locally balanced constraint is presented in Section \ref{sec:61expression}. Finally, Section \ref{sec:concl} concludes the paper.

\section{Preliminaries}\label{sec:pre}

Let $\Sigma=\{0,1\}$. For a sequence $\boldsymbol x=(x_1,x_2,\dots,x_n)\in \Sigma^n$,
its Hamming weight is denoted as $wt(\boldsymbol x)$.
Given integers $\ell$ and $i$, where $1\le i \le n-\ell+1$, a consecutive subword of length $\ell$ starting at the $i$-th coordinate of $\boldsymbol x$ is denoted as $\boldsymbol x[i;\ell]$, i.e., $x[i;\ell]=(x_i,x_{i+1},\dots,x_{i+\ell-1})$.

\begin{definition}
Given a positive even integer $\ell$ and a positive integer $\delta$, a word $\boldsymbol x\in\Sigma^n$ is said to be {\it $(\ell,\delta)$-locally balanced} if for all $1\leq i \leq n-\ell+1$, it holds that
$$\frac{\ell}{2}-\delta \leq wt(\boldsymbol x[i;\ell]) \leq \frac{\ell}{2}+\delta.$$
\end{definition}

\begin{definition}
Given a positive even integer $\ell$ and a positive integer $\delta$, a word $\boldsymbol x\in\Sigma^n$ is said to be {\it strongly $(\ell,\delta)$-locally balanced}, if $\boldsymbol x$ is $(\ell',\delta)$-locally balanced for every even integer $\ell'\geq \ell$.
\end{definition}

\begin{definition}
Let $\Sigma^n(\ell,\delta)$ be the set of all binary $(\ell,\delta)$-locally balanced words of length $n$. Let $\Sigma^n(\ge\ell,\delta)$ be the set of all binary strongly $(\ell,\delta)$-locally balanced words of length $n$. The capacity of such two sets are
\begin{align*}
\mathbb{C}(\ell,\delta) &= \limsup_{n\rightarrow\infty} \frac{\log(|\Sigma^n(\ell,\delta)|)}{n}, \\
\mathbb{C}(\ge \ell,\delta) &= \limsup_{n\rightarrow\infty} \frac{\log(|\Sigma^n(\ge \ell,\delta)|)}{n}.
\end{align*}
\end{definition}

In \cite{Gabrys-ISIT-LBC-20}, Gabrys et al. proposed the definitions above and they aimed at calculating or giving bounds on the values of $\mathbb{C}(\ell,\delta)$ and $\mathbb{C}(\ge \ell,\delta)$. The calculation of $\mathbb{C}(\ell,\delta)$ can be done by analyzing a subgraph $G_{\ell,\delta}$ of the de Bruijn graph of order $\ell$, induced by the $(\ell,\delta)$-locally balanced words of length $\ell$. By the Perron-Frobenius theorem, the value of $\mathbb{C}(\ell,\delta)$ equals $\log \lambda$, where $\lambda$ is the spectral radius of the adjacency matrix of $G_{\ell,\delta}$. The following table of the capacity $\mathbb{C}(\ell,\delta)$ for $4\leq \ell \leq 14$ and $\delta\in \{1,2\}$ is given in \cite{Gabrys-ISIT-LBC-20}.

\begin{table}[!h]
\caption{}
    \centering
    \begin{tabular}{|c||c|c|c|c|c|c|}
    \hline
        $\ell$ & 4 & 6 & 8 & 10 & 12 & 14 \\\hline
        $\delta=1$ & 0.879 & 0.841 & 0.824 & 0.815 & 0.811 & 0.807 \\\hline
        $\delta=2$ & 1 & 0.975 & 0.958 & 0.947 & 0.939 & 0.933 \\\hline
    \end{tabular}
\label{table:capacity}
\end{table}

As for $\mathbb{C}(\ge \ell,\delta)$, we need the following definition.
\begin{definition}
For $\boldsymbol x=(x_1,x_2,\dots,x_n)\in \Sigma^n$, its running digital sum sequence $RDS(\boldsymbol x)=(s_0,s_1,\dots,s_n)\in\mathbb{Z}^{n+1}$ is defined as follows: $s_0=0$, and for $1\leq i \leq n$,
$$s_i=\sum_{j=1}^i (-1)^{1-x_j} = 2wt((x_1,\dots,x_i))-i.$$
Moreover, let $dis(\boldsymbol x)=max_{0\le i\le n}\{s_i\}-min_{0\le i\le n}\{s_i\}$.
\end{definition}

In fact, $s_i$ calculates the difference between the appearances of 1 and the appearances of 0 among the first $i$ coordinates of $\boldsymbol x$ and $dis(\boldsymbol x)$ represents the gap between the largest and smallest integers in $RDS(\boldsymbol x)$. For example, if $\boldsymbol x=(1,0,0,0,0,1)$, then we have $RDS(\boldsymbol x)=(0,1,0,-1,-2,-3,-2)$ and $dis(\boldsymbol x)=1-(-3)=4$.

\begin{definition}
For a given integer $\delta$, a word $\boldsymbol x\in\Sigma^n$ is said to be a $\delta$-RDS word if $dis(\boldsymbol x)\le \delta$. The set of all $\delta$-RDS words of length $n$ is denoted as $\Sigma^n_{RDS}(\delta)$ and the capacity of the set is $$\mathbb{C}_{RDS}(\delta)=\limsup_{n\rightarrow \infty} \frac{\log(|\Sigma^n_{RDS}(\delta)|)}{n}.$$
\end{definition}

The calculation of $\mathbb{C}_{RDS}(\delta)$ is related to a well-known combinatorial problem of counting Dyck paths of bounded height. It has been completely solved in \cite{Gabrys-ISIT-LBC-20} that $\mathbb{C}(\ge \ell,\delta)=\mathbb{C}_{RDS}(2\delta+1)$. In particular, when $\delta=1$, $\mathbb{C}(\ge \ell,1)=\mathbb{C}_{RDS}(3)\approx0.694$.

\section{A coding scheme for strongly locally balanced constraints}\label{sec:SLB}

In this section we consider coding schemes for strongly locally balanced constraints. That is, we want to find an encoding scheme $\mathcal{E}:\Sigma^k\rightarrow \Sigma^n(\ge \ell,\delta)$ with code rate $k/n$ as close to the capacity as possible, together with its corresponding decoding algorithm $\mathcal{D}$. For simplicity, we only consider the case $\ell=4$ and $\delta=1$. The ideas behind the scheme for $\ell=4$ and $\delta=1$ can be naturally generalized to arbitrary $\ell$ and $\delta$.

We want a coding scheme with rate approaching the capacity $\mathbb{C}(\ge 4,1)=0.694$. First, note that by encoding $0$ as $01$ and $1$ as $10$, one can get a trivial encoding scheme $\mathcal{E}:\Sigma^k\rightarrow \Sigma^{2k}(\ge 4,1)$ with code rate $0.5$. We explain our idea by the toy example with code rate $0.6$ as follows.

\begin{example}
We construct $\mathcal{E}:\Sigma^{3k}\rightarrow \Sigma^{5k}(\ge 4,1)$ and thus its rate is $3/5=0.6$. The encoder divides $\boldsymbol{x}\in \Sigma^{3k}$ into $k$ blocks of length 3 and sequentially encodes each block into a string of length $5$. To ensure that the final output $\boldsymbol y$ is strongly $(4,1)$-locally balanced, it suffices to make sure that $dis(\boldsymbol y)\leq 3$. Since $\boldsymbol y$ and $RDS(\boldsymbol y)$ are one-to-one correspondence, we can describe $RDS(\boldsymbol y)=(s_0,s_1,\dots,s_{5k})$ instead of $\boldsymbol y$, and WLOG we fix the range of each entry of $RDS(\boldsymbol y)$ within the interval $[-1,2]$.

Suppose that we are in the position of encoding the block $\{x_{3p+1},x_{3p+2},x_{3p+3}\}$, $0\le p \le k-1$, and the current RDS sequence has $s_{5p}=2$,  i.e., at the top layer of its range. Consider the following 8 diagrams of Dyck paths, which start from layer 2 and are bounded by the interval $[-1,2]$. Each such diagram depicts the trend of the RDS sequence and can be translated into a binary string of length $5$.
\begin{figure}[!h]
    \centering
    \includegraphics[width=0.49\textwidth]{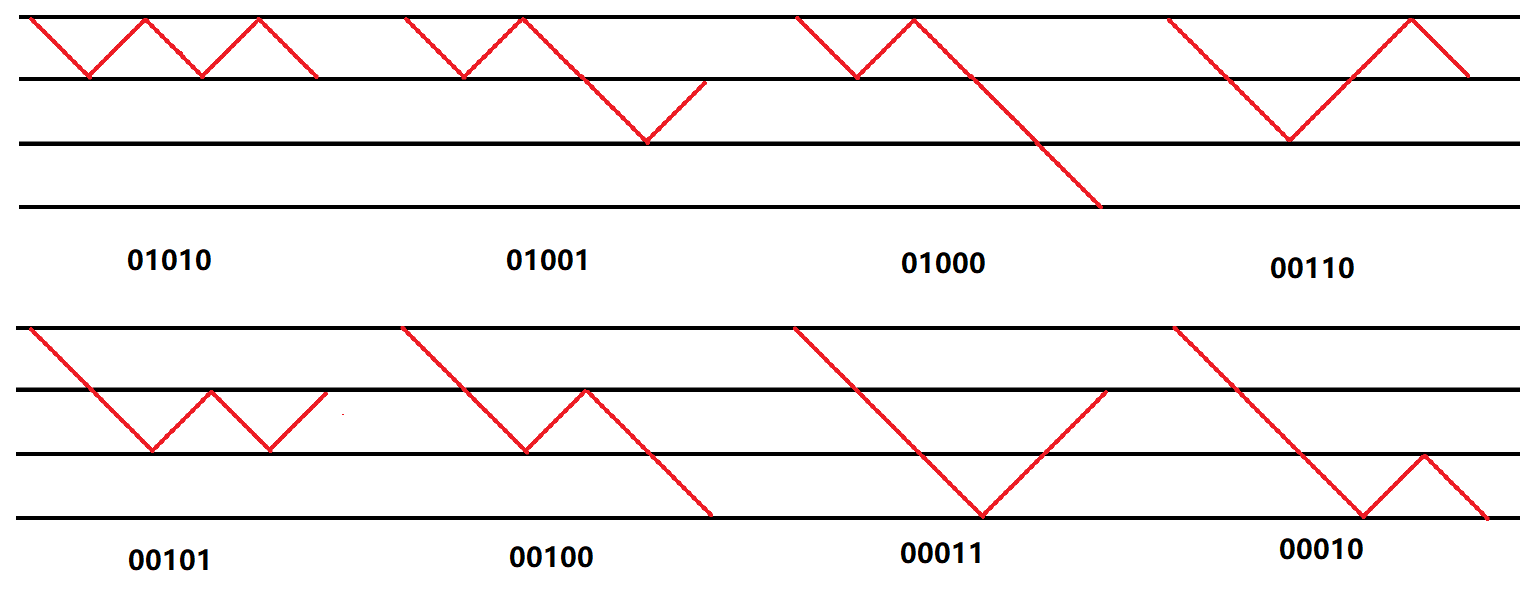}
    \label{Fig:3-5}
\end{figure}

Thus, we may pick an arbitrary one-to-one correspondence mapping between $\Sigma^3$ and these 8 diagrams. Symmetrically, if the current RDS sequence has $s_{5p}=-1$, i.e., at the bottom layer of its range, we also have 8 proper diagrams. It is routine to check that if $s_{5p}\in\{0,1\}$, i.e., the current RDS entry is on the middle two layers, then there are more than 8 diagrams which depict proper trends of the RDS sequence within the interval $[-1,2]$ and we may choose arbitrary 8 of them and build a table-based encoding and decoding algorithm.
\end{example}

Essentially, the scheme above is valid since the number of bounded Dyck paths of length 5 is at least $8=2^3$, no matter in which layer the starting point lies. Motivated by this example, we proceed with the following scheme.

\begin{definition}
Given an integer $m$, let $p(m)$ be the number of Dyck paths bounded by the interval $[-1,2]$ which starts at the layer $2$ (symmetrically,  the layer $-1$) and let $q(m)$ be the number of Dyck paths bounded by the interval $[-1,2]$ which starts at the layer $1$ (symmetrically,  the layer $0$).
\end{definition}

\begin{construction}\label{con:dyckpath}
Given an integer $s$, find the minimum integer $m$ such that $p(m)\ge 2^s$ and $q(m)\ge 2^s$. Then there is an encoding scheme $\mathcal{E}:\Sigma^{sk}\rightarrow \Sigma^{mk}(\ge 4,1)$ for arbitrary $k$ as follows:
\begin{itemize}
    \item Divide $\boldsymbol x\in \Sigma^{sk}$ into $k$ disjoint blocks of size $s$.
    \item Build two tables, each indicating a one-to-one map from $\Sigma^s$ to the two sets of bounded Dyck paths.
    \item Sequentially encode each block of size $s$ by observing the current RDS entry, and then select the corresponding Dyck path from the proper table.
\end{itemize}
\end{construction}

The final output $\boldsymbol y$ is strongly $(4,1)$-locally balanced since $dis(\boldsymbol y)\le 3$ always holds. The decoding algorithm is straightforward based on a table-based search. To maximize the code rate, we need to calculate $p(m)$ and $q(m)$.

\begin{lemma}
For every positive integer $m$, $p(m)=F_{m+1}$ and $q(m)=F_{m+2}$, where $\{F_m\}$ is the Fibonacci sequence with $F_1=1$, $F_2=1$, and $F_k=F_{k-2}+F_{k-1}$ for $k\geq 3$.
\end{lemma}

\begin{IEEEproof}
For a bounded Dyck path starting at layer 2, its next step must go to layer 1, and thus $p(m)=q(m-1)$. For a bounded Dyck path starting at layer 1, its next step goes to either layer 2 or layer 0, and thus $q(m)=p(m-1)+q(m-1)$. Thus we can deduce that $p(m)=q(m-1)=p(m-2)+q(m-2)=p(m-2)+p(m-1)$, with the initial values $p(1)=1$ and $p(2)=2$. Therefore, $p(m)=F_{m+1}$ and $q(m)=p(m+1)=F_{m+2}$.
\end{IEEEproof}

The lemma indicates that $q(m)\geq p(m)$ always holds. Therefore, for every $s$ we only need to find the minimum integer $m$ with $p(m)=F_{m+1}\ge 2^s$ and then we have a coding scheme with rate $s/m$. The rate for small $s$ is summarized in the following table.

\begin{table}[!h]
    \centering
    \begin{tabular}{|c|c|c|c|c|c|c|c|}
        \hline
        s & 2 & 3 & 4 & 5 & 6 & 7 & 8 \\\hline
        m & 4 & 5 & 7 & 8 & 10 & 11 & 13  \\\hline
        s/m & 0.5 & 0.6 & 0.571 & 0.625 & 0.6 & 0.636 & 0.615  \\\hline
        s & 9 &10 & 11 & 12 & 13 & 14 & 15  \\\hline
        m & 14 & 16 & 17 & 18 & 20 & 21 & 23  \\\hline
        s/m & 0.642 & 0.625 & 0.647 & 0.667 & 0.65 & 0.667 & 0.652 \\\hline
    \end{tabular}
\end{table}

When $m$ is sufficiently large, $F_m\approx (\frac{1+\sqrt{5}}{2})^m \approx 2^{0.694m}$, so Construction \ref{con:dyckpath} has rate approaching the capacity $0.694$, as $s\rightarrow \infty$. However, even to achieve rate 0.667 we already need $(s,m)=(12,18)$. Since our map from $\Sigma^s$ to proper Dyck paths is table-based, each table will need to store $2^{12}=4096$ entries, which is rather impractical.

To solve this issue, we move on to a modified construction with rate 0.667 but only requires a table storing 24 entries. The scheme is illustrated by the following state transition diagram.

\begin{figure}[!h]
	\centering
	\begin{tikzpicture}[scale=.68]
     \tikzstyle{edge} = [draw,ultra thick,-stealth,black]
     \node[draw] (A) at (-6,0) {$-1$};
     \node[draw] (B) at (-2,1.5) {$0^+$};
     \node[draw] (C) at (-2,-1.5) {$0^-$};
     \node[draw] (D) at (2,1.5) {$1^+$};
     \node[draw] (E) at (2,-1.5) {$1^-$};
     \node[draw] (F) at (6,0) {$2$};
     \draw[edge] (A) to [bend left] node [above] {110} (B);
     \draw[edge] (A) to node [above] {110,101} (C);
     \draw[edge] (F) to node [below] {001,010} (D);
     \draw[edge] (F) to [bend left] node [below] {001} (E);
     \draw[edge] (B) to  node [below] {100} (A);
     \draw[edge] (B) to [bend left] node [above] {101,110,011} (D);
     \draw[edge] (C) to [bend left] node [below] {010} (A);
     \draw[edge] (C) to [bend left] node [above] {101,110,011} (E);
     \draw[edge] (D) to [bend left] node [above] {011} (F);
     \draw[edge] (D) to [bend left] node [above] {010,001,100} (B);
     \draw[edge] (E) to  node [above] {101} (F);
     \draw[edge] (E) to [bend left] node [above] {010,001,100} (C);
     \draw[edge] (A) .. controls (-4,4.5) and (4,4.5) .. (F) node[pos=0.5,above] {111};
     \draw[edge] (F) .. controls (4,-4.5) and (-4,-4.5) .. (A) node[pos=0.5,below] {000};
\end{tikzpicture}
\end{figure}

\begin{construction}\label{con:modified-dyckpath}
The encoding scheme works as follows. A message $\boldsymbol{x}\in \Sigma^{2k}$ is divided into $k$ blocks of size 2 and we sequentially encode each block into a string of length 3. Initially we have the RDS sequence starting with $s_0=0$, and we set our initial status as $0^+$. When reading the first block $x_1,x_2$, the four choices are set one-to-one correspondence with the four arrows leaving the state $0^+$, which are $\{101,110,011,100\}$. Then recursively in each step we check the current state and then choose one out of the four arrows leaving each state. After encoding all $2k$ information symbols we arrive at a sequence of length $3k$. Finally, we add a `1' if we end in the state $0^+$, $1^+$ and $-1$. Otherwise we add a `0' if we end in the state $0^-$, $1^-$, and $2$. Note that in this way the added bit still guarantees that $s_{3k+1}\in[-2,1]$ and thus does not violate the strongly locally balanced constraint.

The decoding scheme works as follows. Given the encoded string $\boldsymbol{y}$ of length $3k+1$, first check the entry $s_{3k}$ in the RDS sequence. Together with the last bit, we uniquely determine which state we end with. Then the decoding can be done by reading every three bits backwards in a table-based way according to the transition diagram.
\end{construction}

For example, suppose we pick the following table for our coding scheme:

\begin{table}[!h]
    \centering
    \begin{tabular}{|c||c|c|c|c|}
        \hline
        Status & 00 & 01 & 10 & 11 \\\hline
        $-1$  & 110 (to $0^+$) & 110 (to $0^-$) & 101 & 111 \\\hline
        $0^+$ & 101 & 110 & 011 & 100 \\\hline
        $0^-$ & 101 & 110 & 011 & 010 \\\hline
        $1^+$ & 010 & 001 & 100 & 011 \\\hline
        $1^-$ & 010 & 001 & 100 & 101 \\\hline
        $2$   & 000 & 010 & 001 (to $1^+$)& 001 (to $1^-$) \\\hline
    \end{tabular}
\end{table}

Then $\boldsymbol x = (10-01-11-01)$ is encoded as $\boldsymbol y = 011-001-100-110-0$. To decode back, first note that in $RDS(\boldsymbol y)$ we have $s_{12}=0$ and the last bit is 0, so the final state is $0^-$. The edge labelled `110' into the state $0^-$ comes from the state $-1$ and thus $(x_7,x_8)=01$. The edge labelled `100' into the state $-1$ comes from the state $0^+$ and thus $(x_5,x_6)=11$. Repeating this process and then we finally decode $\boldsymbol x = (10-01-11-01)$.

To sum up, we have proved the following:

\begin{theorem}
The coding scheme based on the state transition diagram above is a coding scheme $\mathcal{E}:\Sigma^{2k}\rightarrow\Sigma^{3k+1}(\ge 4,1)$ and thus its rate is $\frac{2k}{3k+1}\rightarrow 0.667$ as $k\rightarrow \infty$.
\end{theorem}

We close this section by discussing the key feature of the state transition diagram. Each state has out-degree 4, which guarantees the encoding process. The incoming arrows into each state have distinct labels, which guarantees the decoding process. In fact,  Construction \ref{con:modified-dyckpath} could be further generalized to $\Sigma^{11k}\rightarrow\Sigma^{16k+1}(\ge 4,1)$ with rate approaching $11/16=0.687$ and $\Sigma^{20k}\rightarrow\Sigma^{29k+1}(\ge 4,1)$ with rate approaching $20/29=0.690$. Compared with Construction \ref{con:dyckpath}, Construction \ref{con:modified-dyckpath} approaches 0.694 faster. Moreover, to achieve the same rate, the table size of Construction \ref{con:modified-dyckpath} will be much smaller than Construction \ref{con:dyckpath}. Due to lack of space, we postpone the details to a future journal version of this paper.

\section{A coding scheme for locally balanced constraints}\label{sec:LB}

Now we consider $(\ell,\delta)$-locally balanced constraints. Note that the two constructions above for strongly $(4,1)$-locally balanced constraints also work for $(\ell,1)$-locally balanced constraints for arbitrary even integer $\ell$. However, the gap between the rate 0.667 and the capacity results as shown in Table \ref{table:capacity} is too large to be negligible. In this section, we attempt to propose a general algorithm to find a coding scheme $\Sigma^{k}\rightarrow\Sigma^{n}(\ell,\delta)$ with rate $k/n$ as close to the capacity as possible.

Given $\ell$, define a directed graph $G_m$ as follows, for all $m\ge \ell-1$. The vertex set of $G_m$ is $\Sigma^m$, the set of all binary strings of length $m$. An arrow from $\boldsymbol x$ to $\boldsymbol y$ exists if and only if their concatenation sequence $\boldsymbol{xy}$ is an $(\ell,\delta)$-locally balanced string of length $2m$. The next algorithm aims at finding the largest integer $s=s(m)$, such that there exists a subgraph of $G_m$ where every vertex has out degree at least $2^s$.

\begin{algorithm}[H]
\caption{}
\hspace*{0.02in} {\bf Input:}
$m$\\
\hspace*{0.02in} {\bf Output:}
$\mathcal{G}_m$, $s$\\
\hspace*{0.02in} {\bf Initially:} set $s:=\lfloor\log\Delta\rfloor$, where $\Delta$  is the maximum degree of $G_m$.
\begin{algorithmic}[1]
\STATE Build the graph $G_m$.
\STATE Delete all vertices with degree less than $2^s$.
\STATE Check the remaining graph. If it is an empty graph then go to Step 4. If it is nonempty and the minimum degree is less than $2^s$, repeat Step 2. Otherwise, go to Step 5.
\STATE Set $s:=s-1$ and go back to Step 1.
\STATE Output the current subgraph as $\mathcal{G}_m$, in which every vertex has degree at least $2^s$. Output the current value of $s$.
\end{algorithmic}
\label{alg:find-subgraph}
\end{algorithm}

Building on this algorithm, a coding scheme for $(\ell,\delta)$-locally balanced constraints is as follows.

\begin{construction}\label{con:graphsearch}
For every $m\ge \ell-1$, implement Algorithm \ref{alg:find-subgraph} to find the corresponding $s(m)$. Compare the values of $s(m)/m$ and find the largest one as $s'/m'$. The output digraph from Algorithm \ref{alg:find-subgraph} with $m'$ as the input is denoted as $\mathcal{G}_{m'}$.

Given any message $\boldsymbol{x}\in\Sigma^{ks'}$ for arbitrary $k$, divide $\boldsymbol{x}$ into $k$ disjoint blocks of size $s'$ and sequentially encode each block of size $s'$ into a string of length $m'$. The encoding starts with a predetermined one-to-one map between $\Sigma^{s'}$ to a subset of the vertex set of $\mathcal{G}_{m'}$. Then, the encoding of the next block is based on a predetermined one-to-one map between $\Sigma^{s'}$ to a subset of the out-going edges from the current vertex. Recursively do the encoding and finally the output is the concatenation of all the strings length $m'$, altogether a string of length $km'$.
\end{construction}

Via computer search, the rate of Construction \ref{con:graphsearch} for $4\leq \ell \leq 14$ and $\delta\in\{1,2\}$ (and the corresponding values of $s'$ and $m'$) is summarized as follows.

\begin{table}[!h]
\center
  \begin{tabular}{|c|c|c|c|}
    \hline
    $\ell$ & 4 & 6 & 8  \\ \hline
    $\delta=1$ & 11/13=0.846 & 12/15=0.8 & 10/13=0.769  \\ \hline
    $\delta=2$ & 1 & 14/15=0.933 & 13/14=0.929  \\\hline
    $\ell$  & 10 & 12 & 14 \\ \hline
    $\delta=1$  & 11/15=0.733 & 11/15=0.733 & 11/15=0.733  \\ \hline
    $\delta=2$  & 8/9=0.889 & 12/14=0.857 & 12/14=0.857  \\\hline
  \end{tabular}
\end{table}

\section{More on the $(6,1)$-locally balanced constraint}\label{sec:61expression}

In this section we derive the linear recurrence relation on the size of $\Sigma^n(6,1)$. There are mainly two motivations for analyzing this formula.

Firstly, while theoretically computing the capacity $\mathbb{C}(\ell,\delta)$ can be done by the spectral graph theory approach as explained in \cite{Gabrys-ISIT-LBC-20}, the exponentially growing size of the adjacency matrix makes all known algorithms of spectral radius impractical. This is also why Table \ref{table:capacity} ends with $\ell=14$. For general $\ell$ and $\delta$, determining $\mathbb{C}(\ell,\delta)$ seems to be a very difficult problem. We want to try to find the linear recurrence relation and thus the exact formula on the size of $\Sigma^n(\ell,\delta)$, or try to find linear recurrence inequalities which can provide upper or lower bounds. The formula for the size of $\Sigma^n(4,1)$ can be easily solved and thus the next step starts with the case $\Sigma^n(6,1)$.

Secondly, an explicit formula might lead to a coding scheme based on recursive enumeration techniques, based on a partial order of the set of proper codewords, which is indeed the case for coding schemes under run-length limited constraints, see for example \cite{Nguyen-IT-constrainedcodes-21}. To be honest, it seems that our formula below for $\Sigma^n(6,1)$ does not lead to an explicit coding scheme at this moment, but we believe that it might shed light on some new strategies for coding schemes other than Construction \ref{con:graphsearch}.

Now we present the main result of this section.

\begin{theorem}\label{thm:recurrence}
Let $f_n$ be the size of $\Sigma^n(6,1)$, the set of $(6,1)$-locally balanced binary words of length $n$. Then we have the following linear recurrence relation
$$f_{n+12}=f_{n+11}+f_{n+10}+f_{n+9}-f_{n+6}-f_{n+4}-f_{n+3}+f_{n}.$$
\end{theorem}

The proof of this theorem is broken into several lemmas. For any binary string $\boldsymbol z$ let $f_n(\boldsymbol z)$ be the number of words in $\Sigma^n(6,1)$ with $\boldsymbol z$ as a prefix.
Given integers $0\leq t \leq s$,  let $X_n(6,1;s,t)$ be the words $\boldsymbol x\in\Sigma^n(6,1)$ such that $wt(\boldsymbol x[1;s])=t$ and let $f_{n}(s,t)=|X_n(6,1;s,t)|$.

\begin{lemma}\label{lemma1}
$f_n = f_{n+3}(000) + f_{n+3}(111) = f_{n+4}(1000) + f_{n+4}(0111)$.
\end{lemma}

\begin{IEEEproof}
For any $\boldsymbol x\in\Sigma^n(6,1)$, consider the weight of its first three entries and then $\Sigma^n(6,1)$ is a disjoint partition of $X_n(6,1;3,0), X_n(6,1;3,1), X_n(6,1;3,2), X_n(6,1;3,3)$ and thereby we have $f_n=f_n(3,0)+f_n(3,1)+f_n(3,2)+f_n(3,3)$.

For any $\boldsymbol x\in X_n(6,1;3,0)$, $\boldsymbol x=(0,0,0,x_4,x_5,x_6,\dots,x_n)$ and $\{x_4,x_5,x_6\}$ must contain at least two 1's. Let $\boldsymbol{x}'\triangleq111\boldsymbol{x}$. Then we have $wt(\boldsymbol{x}'[1;6])=3$, $wt (\boldsymbol{x}'[2;6])\in\{2,3\}$ and $wt (\boldsymbol{x}'[3;6])\in\{2,3\}$. Moreover, $\boldsymbol{x}'[i;6] = \boldsymbol{x}[i-3;6]$ for $i\ge 4$ and is thereby $(6,1)$-locally balanced. Thus we have $\boldsymbol{x}'\in \Sigma^{n+3}(6,1)$. Similarly, it is also routine to check that for any $\boldsymbol x\in X_n(6,1;3,1)$, it also holds that $111\boldsymbol{x}\in \Sigma^{n+3}(6,1)$. Meanwhile, for any sequence $111\boldsymbol{x}\in\Sigma^{n+3}(6,1)$, delete the prefix $111$ and the remaining $\boldsymbol x$ must satisfy $\boldsymbol x\in\Sigma^n(6,1)$ and its first three entries has weight either 2 or 3. To sum up, we have proved $f_{n+3}(111)=f_n(3,0)+f_n(3,1)$. Symmetrically we have $f_{n+3}(000)=f_n(3,2)+f_n(3,3)$ and thus $f_n = f_{n+3}(000) + f_{n+3}(111)$.
The proof of $f_n = f_{n+4}(1000) + f_{n+4}(0111)$ is similar and thus omitted.
\end{IEEEproof}

\begin{lemma}\label{lemma2}
$f_{n+2}(110) = f_n (0) - f_n (0111)$ and $f_{n+2}(001) = f_n(1)-f_n(1000)$.
\end{lemma}

\begin{IEEEproof}
For any $\boldsymbol x\in \Sigma^n(6,1)$ starting with 0, $11\boldsymbol x\notin \Sigma^{n+2}(6,1)$ if and only if $\boldsymbol x$ starts with $0111$. Thereby $f_{n+2}(110) = f_n (0) - f_n (0111)$. Similarly we have $f_{n+2}(001) = f_n(1)-f_n(1000)$.
\end{IEEEproof}

\begin{lemma}\label{lemma3}
$f_{n+4}(0000)=f_n(11)$ and $f_{n+4}(1111)=f_n(00)$.
\end{lemma}

\begin{IEEEproof}
Any $\boldsymbol x\in \Sigma^{n+4}(6,1)$ starting with 0000 must start with 000011. For any $\boldsymbol x\in \Sigma^{n}(6,1)$ starting with $11$, we can add the prefix 0000 and the resultant string is still $(6,1)$-locally balanced. Therefore $f_{n+4}(0000)=f_n(11)$. Similarly we have $f_{n+4}(1111)=f_n(00)$.
\end{IEEEproof}

\begin{lemma}\label{lemma4}
$f_{n+3}-f_{n+2}-f_{n+1}-f_{n} = -f_{n+1}(0000)-f_{n+1}(1111)-f_n(000)-f_n(111)$.
\end{lemma}

\begin{IEEEproof}
First consider the difference $f_{n+3}-f_{n+2}$. For any word $\boldsymbol x\in\Sigma^{n+2}(6,1)$, $0\boldsymbol x \in\Sigma^{n+3}(6,1)$ if and only if the first five entries of $\boldsymbol x$ has weight $2,3,4$, and $1\boldsymbol x \in\Sigma^{n+3}(6,1)$ if and only if the first five entries of $\boldsymbol x$ has weight $1,2,3$. Therefore, $f_{n+3}-f_{n+2}=f_{n+2}(5,2)+f_{n+2}(5,3)$.

Next, for $\boldsymbol x \in X_{n+2}(6,1;5,2)$, by puncturing the first entry we get a string in $\boldsymbol y \in X_{n+1}(6,1;4,1) \cup X_{n+1}(6,1;4,2)$. On the other hand, for any $\boldsymbol y \in X_{n+1}(6,1;4,1) \cup X_{n+1}(6,1;4,2)$ there is a unique way to add a bit in the beginning to get $\boldsymbol x \in X_{n+2}(6,1;5,2)$. Therefore, $f_{n+2}(5,2) = f_{n+1}(4,1) + f_{n+1}(4,2)$. Similarly, $f_{n+2}(5,3) = f_{n+1}(4,3) + f_{n+1}(4,2)$. Following a similar analysis, we also have $f_{n+1}(4,2)=f_{n}(3,1)+f_n(3,2)$.

Thus we have the following computation:
\begin{align*}
&f_{n+3}-f_{n+2}  -f_{n+1}-f_{n}  \\
&= f_{n+2}(5,2)+f_{n+2}(5,3)-f_{n+1}-f_{n} \\
&= f_{n+1}(4,1) + 2f_{n+1}(4,2) + f_{n+1}(4,3) -f_{n+1}-f_{n} \\
&= -f_{n+1}(0000)-f_{n+1}(1111) + f_{n}(3,1)+f_n(3,2)-f_{n} \\
&= -f_{n+1}(0000)-f_{n+1}(1111)-f_n(000)-f_n(111).
\end{align*}
Thus the lemma follows.
\end{IEEEproof}

With these preparations, now we are ready for the proof of the linear recurrence relation.

\begin{IEEEproof}[Proof of Theorem \ref{thm:recurrence}]
We have the following deduction.
\begin{align}
		&f_{n+4}(0111) + f_{n+4}(1000) = f_n \\
		\Rightarrow &f_{n+6}(110) + f_{n+6}(001) = f_{n+4} - f_n \\
		\Rightarrow &f_{n+6}(110) + f_{n+6}(001) + f_{n+6}(111) + f_{n+6}(000) \nonumber\\
		&= f_{n+4} + f_{n+3} - f_n  \\
		\Rightarrow &f_{n+9}(000) + f_{n+9}(111) \nonumber\\
		&= f_{n+6}(01) + f_{n+6}(10) + f_{n+4} + f_{n+3} - f_n \\
		\Rightarrow &f_{n+10}(0000)+f_{n+10}(1111)+f_{n+9}(000)+f_{n+9}(111)\nonumber\\
		&= f_{n+6} + f_{n+4} + f_{n+3} - f_n
	\end{align}

Equation (1) follows Lemma \ref{lemma1}.

Equation (2) follows Lemma \ref{lemma1} and Lemma \ref{lemma2}, which implies that $f_{n+6}(110) + f_{n+6}(001) = f_{n+4}(1) + f_{n+4}(0) - f_{n+4}(1000) - f_{n+4}(0111) = f_{n+4} - f_{n}$.

Equation (3) follows Lemma \ref{lemma1}, which implies that $f_{n+6}(000) + f_{n+6}(111) = f_{n+3}$.

Equation (4) follows Lemma \ref{lemma1} and Equation (3), which implies that $f_{n+9}(000) + f_{n+9}(111) = f_{n+6} = f_{n+6}(10) + f_{n+6}(01) + f_{n+6}(11) + f_{n+6}(00) = f_{n+6}(10) + f_{n+6}(01) + f_{n+4} + f_{n+3} - f_{n}$.

Equation (5) follows Lemma \ref{lemma3}, which implies that $f_{n+10}(0000) + f_{n+10}(1111) = f_{n+6}(11) + f_{n+6}(00)$.

Finally, using Equation (5) and Lemma \ref{lemma4}, we can deduce that $f_{n+12}-f_{n+11}-f_{n+10}-f_{n+9}=-f_{n+10}(0000)-f_{n+10}(1111)-f_{n+9}(000)-f_{n+9}(111)=-f_{n+6} - f_{n+4} - f_{n+3} + f_n $, and thus the recurrence relation holds.
\end{IEEEproof}

\medskip

Note that by solving the recurrence relation, one can find that $f(n)\approx 1.791^n$ and thus $\mathbb{C}(6,1)=0.841$, which is the same as the result from the spectral graph theory method as expected.

A final remark is that unfortunately the recurrence relation does not trivially lead to a coding scheme based on  enumerations as in the case of run-length limited constrained codes. However, we believe that the exact formula might shed light on some new coding strategies.  Moreover, finding the recurrence relation (or proper forms of inequalities) on the size of $\Sigma^n(\ell,\delta)$ for general $(\ell,\delta)$ is an interesting and challenging direction.

\section{Conclusion}\label{sec:concl}

In this paper we further study the locally balanced constraints proposed in \cite{Gabrys-ISIT-LBC-20}. We propose two coding schemes for strongly locally balanced constraints. In particular, for the strongly $(4,1)$-locally balanced constraint we have a scheme with rate $0.667$ close to the capacity limit $0.694$, based on a simple look-up table. For locally balanced constraints we propose an algorithm to find a table-based coding scheme, with rate as close to the capacity limit as possible. Moreover, we give an additional result on the linear recurrence relation on the size of $\Sigma^n(6,1)$, and new coding strategies based on such formulas are considered for future research.

\pagebreak

\end{document}